\documentclass[11pt]{article}

\usepackage{fullpage}
\usepackage{amsmath}
\usepackage{graphicx}
\usepackage{authblk}
\usepackage[center]{caption}
\usepackage{refstyle}
\usepackage[breaklinks=true,colorlinks=true,linkcolor=blue,urlcolor=blue,citecolor=blue]{hyperref}

\begin{document}
\title{\textbf{Surface scaling behaviour of size-selected Ag-nanocluster film growing under subsequent shadowing process} }
\author[1,2]{Pintu Barman}
\author[1,2]{Anindita Deka}
\author[1,3,4]{Shyamal Mondal}
\author[1,5]{Debasree Chowdhury}
\author[1]{Satyaranjan Bhattacharyya}
\affil[1]{Saha Institute of Nuclear Physics, 1/AF Bidhan Nagar, Kolkata-700064, India}
\affil[2]{Homi Bhaba National Institute, Training School Complex, Anushakti Nagar, Mumbai 400094, India}
\affil[3]{Maharaja Manindra Chandra College, 20 Ramkanto Bose Street, Kolkata-700003, India}
\affil[4] {Center for Materials and Microsystems, Fondazione Bruno Kessler, via Sommarive 18, Trento-38123, Italy}
\affil[5] {Dipartimento di Fisica, Università di Genova, via Dodecaneso 33, 16146 Genova, Italy}
\date{}
\maketitle
\abstract
Surface morphology of size-selected silver nanocluster films grown by dc magnetron sputtering has been investigated by means of an atomic force microscopy (AFM). From the height-height correlation functions ( HHCF) obtained from corresponding AFM images, the scaling exponents are calculated and two types of growth regimes have been observed. In the first regime, the growth exponent is  found to be $\beta_1=0.27\pm0.07$ close to the KPZ growth exponent, while in the second growth regime shadowing effect plays dominant role which gives the  growth exponent value $\beta_2=0.88\pm0.28$. On the other hand for the whole deposition regime, the roughness exponent value is found to be constant around $\alpha=0.76\pm0.02$. UV-vis spectroscopy measurement suggests how the average reflectance of the film surface changes with different growth times.
\begin{figure}[htb]
\centering
\captionsetup{width=.8\linewidth}
\includegraphics[scale=.55]{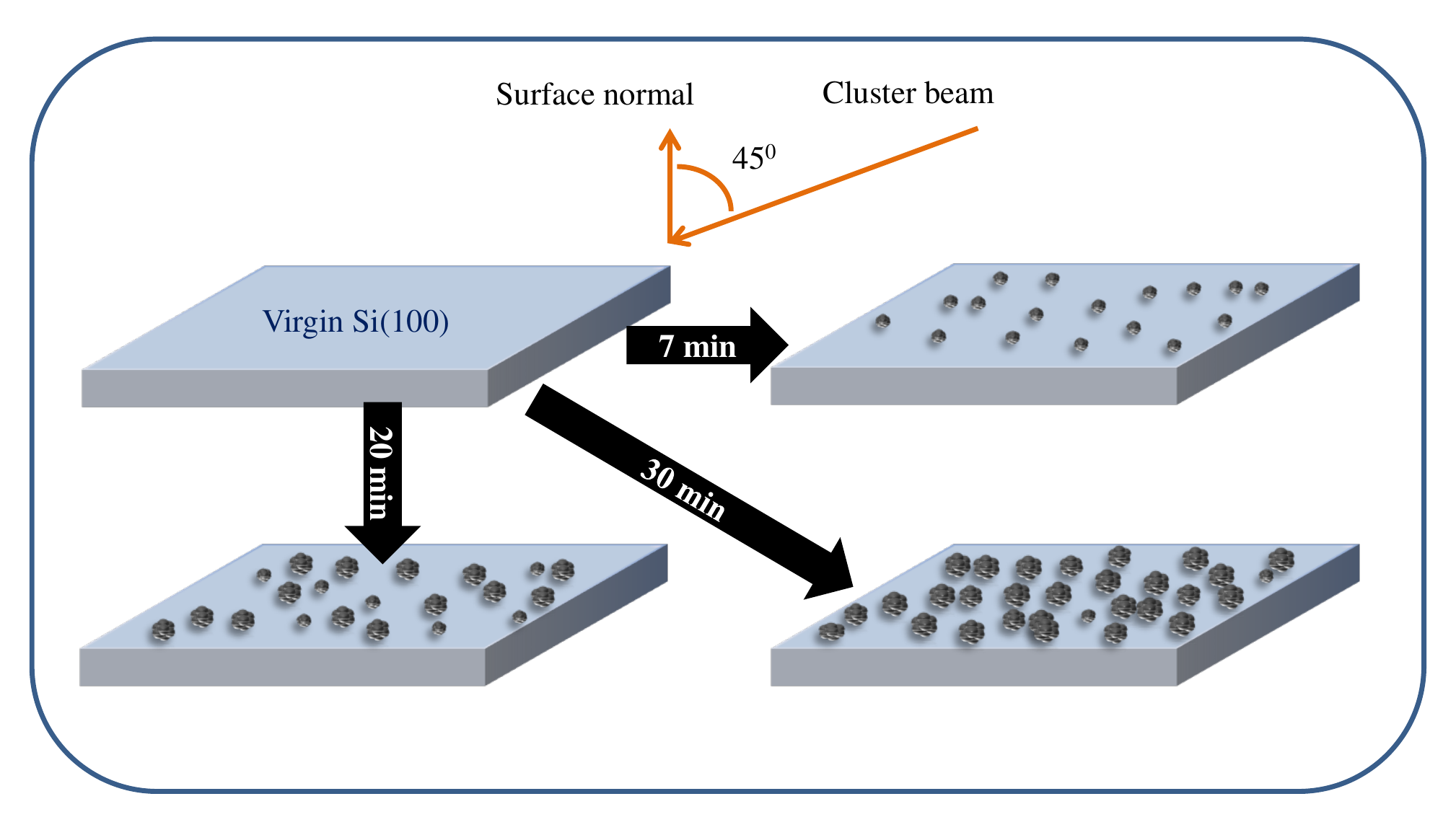}
\label{fig:Scheme}
\end{figure}

\section{Introduction}
The study of kinetic roughening of growing surface under non equilibrium conditions has been a very active field for many years \cite{ref1,ref2,ref3,ref4}. In the field of materials science, formation of thin film by deposition of nanoclusters has become an interesting topic from fundamental as well as application point of view. For low energy deposition or soft-landing deposition to occur, cluster kinetic energy must be very low. For a growing surface morphology due to low energy cluster beam deposition, there are mainly three parameters which play crucial roles and these are deposition, desorption and surface diffusion of clusters. The parameter desorption is mainly dependent on the kinetic energy of the cluster beam and sticking coefficient, whereas the\begin{figure}[htb]
\centering
\captionsetup{width=.8\linewidth}
\includegraphics[scale=.8]{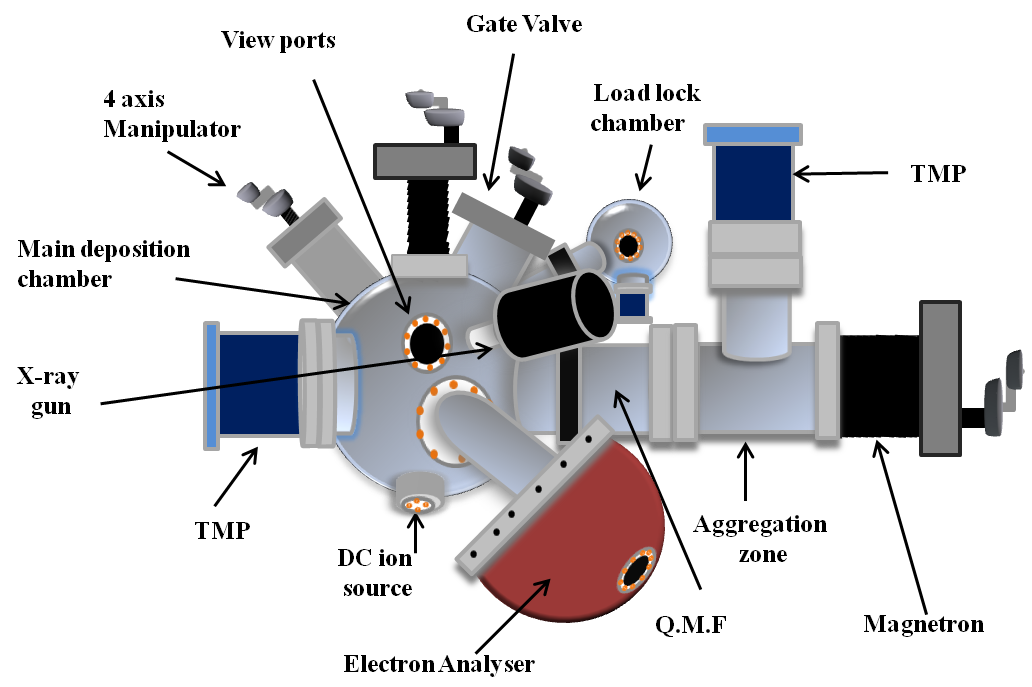}
\caption{\small Schematic diagram of Nanocluster deposition system}
\label{fig:fig_1}
\end{figure}
surface diffusion term mainly depends on the temperature of the substrate surface as well as the size and structure of the deposited cluster \cite{ref5, ref6}. By balancing these three parameters a self-affine scaling behavior of the growing surface can be observed \cite{ref7}. A detailed study on the growth dynamics of the surface morphology is possible by characterizing the obtained scaling exponents under the framework of dynamic scaling theory \cite{ref8}. These scaling exponents are basically roughness exponents $\alpha$, which denotes the degree of surface irregularity for short length scales and growth exponent $\beta$, which signifies the change of interface width over time. In dynamic scaling theory, depending on certain values of $\alpha$ and $\beta$, some universality classes exist, so that most of the experimental systems could be represented by any of these classes \cite{ref9,ref10,ref11}. However, for many experimental systems, it has been observed that the obtained scaling exponents do not fall into any of the pre-existing models.  Growth dynamics of cluster assembled  thin films have been reported by many groups, but the study of dynamic scaling of thin films prepared by size-selected clusters are not much explored. Previously we have studied the growth dynamic behavior of size-selected Cu nanoclusters on Si substrates \cite{ref12}. In this work, we present the characterization of the morphology of the surface grown by size selected silver clusters by varying the deposition time on Si (100) substrates. We choose to use Silver as our depositing cluster materials because of its immense applications in catalytic\cite{ref12a},  biomedical\cite{ref12b}, electronics devices\cite{ref12c},  photocatalytic\cite{ref12d}, solar cell\cite{ref12e} etc. Because of their high surface to volume ratio in nanoscale regime, Ag nanoclusters have some unique properties from its bulk counterpart. In many applications size of the deposited Ag nanoparticles/nanoclusters plays a crucial role such as in some chemical reactions where clusters are used as catalyst, a small change in cluster size leads to significant influence on the activity of the reactions\cite{ref12f,ref12g}. Moreover, size selected clusters can be used in glucose sensors to enhance the sensitivity\cite{ref12h}. Also silver nanoparticles have a very unique optical response as surface plasmon resonance and found very promising in memory device applications\cite{ref12i}. For this purpose control over the size on the synthesis technique is very essential for further development. Using our nanocluster deposition facility it is quite possible to control the size of deposited Ag-nanoclusters depending on the requirements. We also show how the shadowing effect plays an indispensable part in the evolution of surface growth.

\section{Dynamic scaling theory}
Dynamic scaling theory provides very useful concepts in determining the surface morphology and understanding the growth behavior of thin film surface. The roughness of a film surface can be characterized statistically by determining the interface width $w(r,t)$, which describes the fluctuations of surface heights $h(\vec{r},t)$ around an average surface height $\left<h\right>$. The interface width is defined as $w(r,t)=\left<[ h(\vec{r},t)-<h> ]^2\right>^{1/2}_{\vec{r}}$, where $t$ is the deposition time and   $\left<...\right>_{\vec{r}}$ denotes the average over all $\vec{r}$   in a system of size $L(r\leq L)$. From the expression of interface width, two important quantities are derived viz. local width where r$\ll$L and global width where r=L. According to dynamic scaling theory, both these interface widths follow the Family-Vicsek relation \cite{ref13} which is given as,
$$w(r,t)=t^{\alpha/z}f(r/t^{1/z})$$
where $f(u)$ being the scaling function and it behaves as,

\begin{equation*}
f(u)\propto
\begin{cases}
u^\alpha,\hspace{.7cm} if \hspace{.5cm} u\ll1
\\
const,\hspace{.3cm} if\hspace{.5cm} u\gg1
\end{cases}
\end{equation*}

Here the exponent $\alpha$ is called the roughness exponent which describes the lateral correlation of the surface roughness and the exponent $z$ is called the dynamic exponent which describes how the lateral correlation length of the surface scales with time. The scaling function here constitutes two different scaling regimes depending on its argument $u\equiv (r/t^{1/z})$ . For small $u$, the the interface width $w(r,t)$ varies as $r^\alpha$. Whereas for $u\gg1$, the the interface width $w(r,t)$ increases as well, and the dependence also follows a power law, i.e. $w(r,t)\propto t^\beta$ . Here $\beta=\alpha/z$ is called the growth exponent which characterizes the time dependent dynamics of the surface roughening process \cite{ref8}. These two regions are separated by a ‘crossover length’ at $ r=ξ\xi$ called lateral correlation length which depends on the deposition time as $t^{1/z}$. Within the lateral correlation length $\xi$, the heights between two points are considered to be correlated. 

Apart from the direct computation of the interface width, there are other functions in order to investigate the scaling properties of the thin film surface which can be calculated statistically. One of the function is the height-height correlation function (HHCF) calculated in direct space, which measures the lateral auto correlation of the surface height. The HHCF for a homogeneous, isotropic random surface is given as,$$G(r,t)=\left<[h(\vec{r_2},t)-h(\vec{r_1},t]^2)\right>\ , \hspace{.5cm}r=|{\vec{r_2}-\vec{r_1}}|$$
\\
where $h(\vec{r_1},t)$and $h(\vec{r_2},t)$ are the surface heights at position $\vec{r_2}$ and $\vec{r_1}$ respectively at deposition time $t$ \cite{ref14}.\\
For self affine surface, the time dependent HHCF satisfies the relation,$$G(r,t)=2[w(t)]^2f(r/\xi)$$
with dynamic scaling requirement of $z=\alpha/\beta$

Thus, HHCF satisfying the above condition shows asymptotic behavior and is given by \cite{ref8},

\begin{equation*}
G(r,t)\propto
\begin{cases}
r^{2\alpha}\hspace{.5cm} if\hspace{.5cm} r\ll\xi
\\
2w^2 \hspace{.401cm}if\hspace{.5cm} r\gg\xi
\end{cases}
\end{equation*}
The other function which manifests the scaling behavior of random surface is the power spectral density (PSD) function which is the Fourier transform of surface height measured in reciprocal space. The PSD function helps to reveal the periodic surface features that might otherwise appear random by representing the amplitude of surface’s roughness as a function of spatial frequency of roughness. It is defined as,$$PSD(k,t)=\left< H(k,t)H(-k,t)\right>$$
where $H(k,t)$ is the Fourier transform of surface height in a system of size $L$ and $k$ is the spatial frequency in reciprocal space. Statistically, PSD function also follows the Family-Vicsek relation and can be expressed in terms of scaling function $s(kt^{1/z})$ as;
\begin{equation*}
PSD(k,t)= k^{-(2\alpha+1)}s(kt^{1/z})
\end{equation*}
where the generalized form of scaling function is given by\cite{ref15,ref16},
\begin{equation*}
s(kt^{1/z})\propto
\begin{cases}
({kt^{1/z}})^{2\alpha+1}\hspace{1.2cm};(kt^1/z)\ll1
\\
Const\hspace{2cm};(kt^1/z)\gg1
\end{cases}
\end{equation*}

\section{Experimental procedure}
In this work, we performed  deposition of size-selected silver nanoclusters on Si(100) substrates containing native oxide, utilizing a magnetron-based gas aggregation type nanocluster source (Model: NC200U, OAR, UK). The schematic diagram of the experimental set up is shown in \Figref{fig_1}. Before deposition, silicon substrates were properly cleaned in acetone and propanol to remove the organic contamination. To produce size-selected metal nanocluster, a silver target with 99.99\% purity is used as a target of the magnetron and sputtered by Ar gas to generate the plasma. After sputtering the target by Ar$^{+}$ ion, the sputtered Ag atoms were condensed into clusters of various sizes with the help of buffer gas inside an aggregation chamber in a vacuum environment and then the clusters were swept away by a gas stream from the aggregation chamber into the next chamber through a nozzle and collimated into a cluster beam. In the next chamber, the collimated cluster beam was passed through a Quadrupole Mass Filter (QMF). In QMF a particular mass of clusters from all possible masses of formed clusters are allowed to pass. In this experiment clusters of 4 nm diameter are selected, with a mass $m=(4/3)\pi  r^{3}\rho$, where $r$ is radius of cluster and $\rho$ is the density of the cluster material. The number of atoms in a cluster of radius $r$ is given by $n=(r/r_w)^3$ where $r_w$ is the Wigner-Seitz radius. Using this equation, the number of atoms for a silver cluster with  diameter 4 nm is found to be $n$=1750, with $r_w$=0.166 nm for silver. Size-selected cluster beam was then directed towards the main deposition chamber where the substrate was placed vertically by an angle of $45^o$ with the cluster beam. A detail elaboration of production of size selected clusters inside the nanocluster system has been reported by Mondal and Bhattacharyya \cite{ref17}. In this experiment size-selected, Ag nanoclusters with 4 nm diameter  were used to be deposited on $10 \times10$ $mm^2$ Si (100) substrates. The base pressure inside the main deposition chamber and the aggregation chamber were $2\times10^{-9}$ mbar \&  $5\times10^{-4}$ mbar respectively and during deposition, pressure rises inside the main chamber and the aggregation chamber to $4.8\times10^{-4}$ mbar and $2.5\times10^{-1}$ mbar respectively. A series of samples were prepared for different deposition time viz., 7 min, 10 min, 15 min, 20 min, 25 min and 30 min. Morphological characterization of the size-selected Ag nanocluster films was carried out by an atomic force microscope in tapping mode (Model: Nanoscope IV, make: Veeco, USA), Scanning Electron Microscope, x-ray photoelectron spectroscopy ( Model: Class150 Bolt-on, Make: VSW, UK) facility attached with the nanocluster deposition unit. The optical properties were measured from the UV-vis reflectence spectra of the samples which are recorded using PerkinElmer lambda 750 spectrophotometer.

\section{Experimental Result and Discussions}

\begin{figure}[htb]
\centering
\captionsetup{width=.8\linewidth}
\includegraphics[scale=.58]{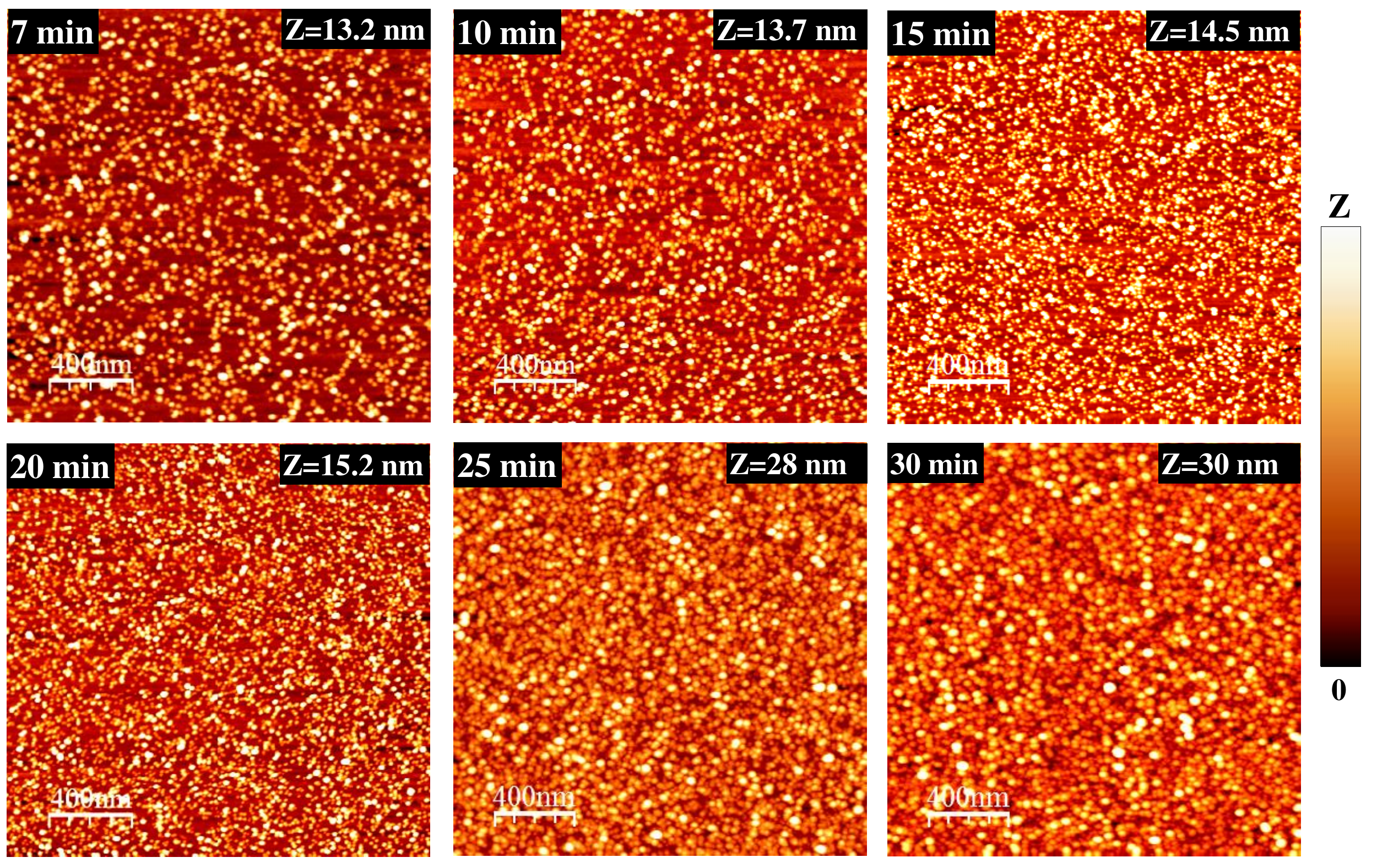}
\caption{\small Typical $(2\mu m\times2\mu m)$ AFM images of size-selected Ag nanocluster film deposited on Si (100) substrates for 7 min, 10 min,15 min, 20 min, 25 min and 30 min}
\label{fig:fig_2}
\end{figure}

\Figref{fig_2} shows the AFM images representing morphology of the Ag nanocluster films deposited on Si (100) substrates for all the deposition time. From the images it is seen that the surface morphology formed by Ag clusters are increases with deposition time. From the corresponding AFM images, it is clearly observed that height of the Ag
nanoclustes does not increase much for 7 min, 10 min,15 min and 20 min of deposition time, only density increases. While for 25 min and 30 min of deposition times, both height and density of the Ag nanocluster film is found to be very high compared to previous deposition times.

\begin{figure}[htb]
\centering
\captionsetup{width=.8\linewidth}
\includegraphics[scale=.55]{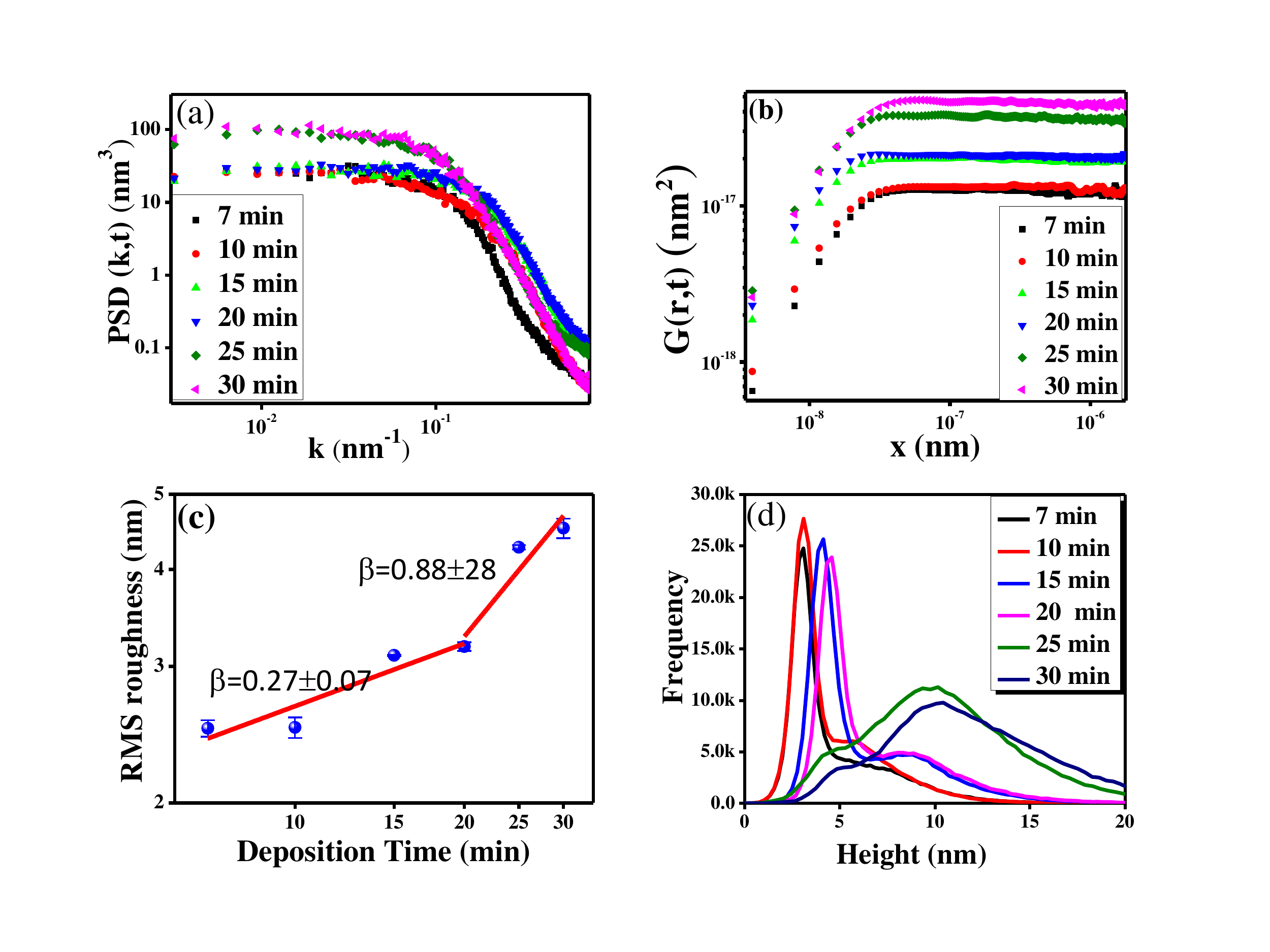}
\caption{\small (a) Log-log plot of the Power spectral density (PSD) functions vs wave number k for different deposition time, (b) Logarithmic plot of height-height correlation function G(r,t) vs lateral distance r, (c) RMS roughness fordifferent deposition time, (d) Height distriburion of Ag nanocluster film for different deposition time}
\label{fig:fig_3}
\end{figure}

From the Power Spectral Density (PSD) curve we can determine the kinetic roughening of the films. The log-log plot of PSD (k,t) versus k for all deposition time is shown in \Figref{fig_3}(a). All the PSD curves give two types of variations. Initially for low frequencies (for small k value) the curves are almost saturated which indicates the absence of any kind of lateral correlation in the surface roughness within these range. And for high frequencies (for large k value) all the PSD curves show a linear behavior with negative slope which indicates the existence of kinetic roughening within that particular length scales. Also the non-occurrence of any kind of peak on the PSD curves reveals the absence of surface periodicity in the growth process. From the saturation region of the PSD curve, it is observed that upto 20 min deposition time all curves overlap and after 20 min of deposition time, a sharp upward shift takes place and then again there is overlapping of the curves. From this type of behavior of the PSD curves, we can expect a transition in the growth of nanoclusters at the deposition time of 20 min.

\Figref{fig_3}(b) shows the logarithmic plot of height-height correlation function G(r,t) vs scanning length r for all the samples deposited for various time scales. For all deposition time, it is seen that the HHCF initially shows a linear increase for small value of r and gets saturated for larger value of r.  It is also observed that there is an upward shift of the graphs with the increase of deposition time for the complete range of r. 
Since HHCF does not show any oscillatory behavior for the entire range along x axis, it confirms that deposited surfaces are self-affine. Thus from the slope of the G(r,t) functions for small r value, we can determine the roughness exponent value ($\alpha$) by using the equation $G(r,t)= r^{2\alpha}$. It is found that for all deposition time, the roughness exponent is almost constant and equals to $\alpha=0.76\pm0.02$. Also from the saturation region of the HHCF graph, we can calculate the root mean-square (RMS) roughness ($w$) of the films using the formula $w=\sqrt{G(r,t)/2}$ . The variation of RMS roughness with deposition time is shown in \Figref{fig_3}(c) and from the graph, two types of variation of the growth exponent value is clearly seen, initially the RMS roughness increases very slowly with $\beta$ value $0.26\pm0.07$ for deposition time 7 min to 20 min and after that the RMS roughness increases sharply with another $\beta$ value $0.88\pm0.28$ for deposition time 20 min to 30 min respectively.

When we plot the height distribution function for each sample from the corresponding AFM images, we found that in all cases height distribution is bimodal in nature, but the intensity falls down after a certain time of deposition as shown in \Figref{fig_3}(d). This can be explained by considering the variation of number density with deposition time. Initially number of Ag cluster on the surface is very low, so during scanning the surface, AFM tip can detect both the surface as well as large number of clusters distributed on the surface of the substrate. With the increase of deposition time, initially the number density also increases, but after a certain time of deposition the number density starts decreasing, this may be due to the diffusion of clusters on the surface and formation of cluster islands. Due to this possibility, the surface of higher deposition time exhibits low intensity peaks in bimodal height distribution.

\begin{figure}[htb]
\centering
\captionsetup{width=.8\linewidth}
\includegraphics[scale=1]{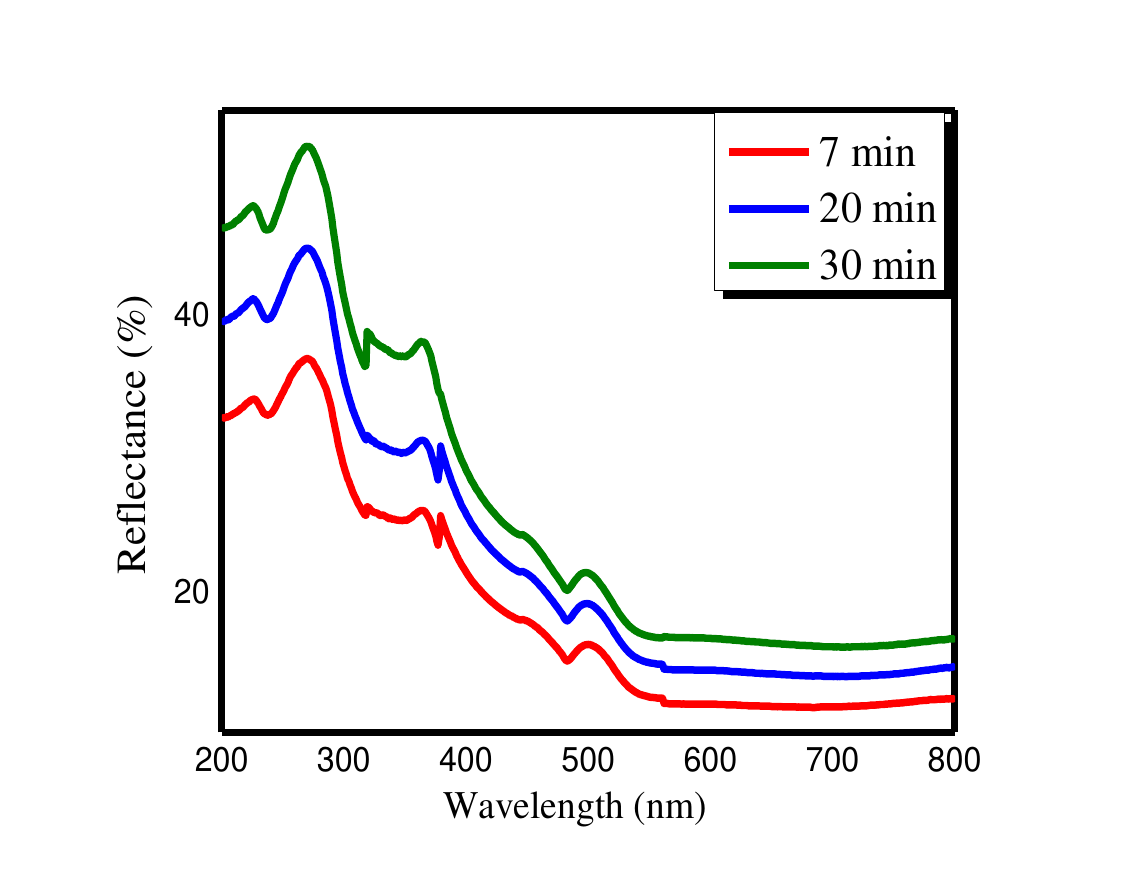}
\caption{\small UV-Vis reflectance spectra of Ag nanoclusters deposited Si(100) for 7 min, 20 min, 30 min}
\label{fig:fig_4}
\end{figure}

In addition we investigate the UV-vis reflectance spectra of size selected Ag nanoclusters deposited on Si (100) for different time is shown in \Figref{fig_4}. From the plot, it is found that the average reflectance increases with deposition time and the values of average reflectance for 7min, 20 min and 30 min are 19\%, 22\% and 26\% respectively. With the increase of deposition time surface coverage increases which means more silver nanoclusters are deposited on the substrate. As a result of this, with the increased Ag nanoclusters on the substrates a higher surface reflectance has been found.

\begin{figure}[htb]
\centering
\captionsetup{width=.85\linewidth}
\includegraphics[scale=.75]{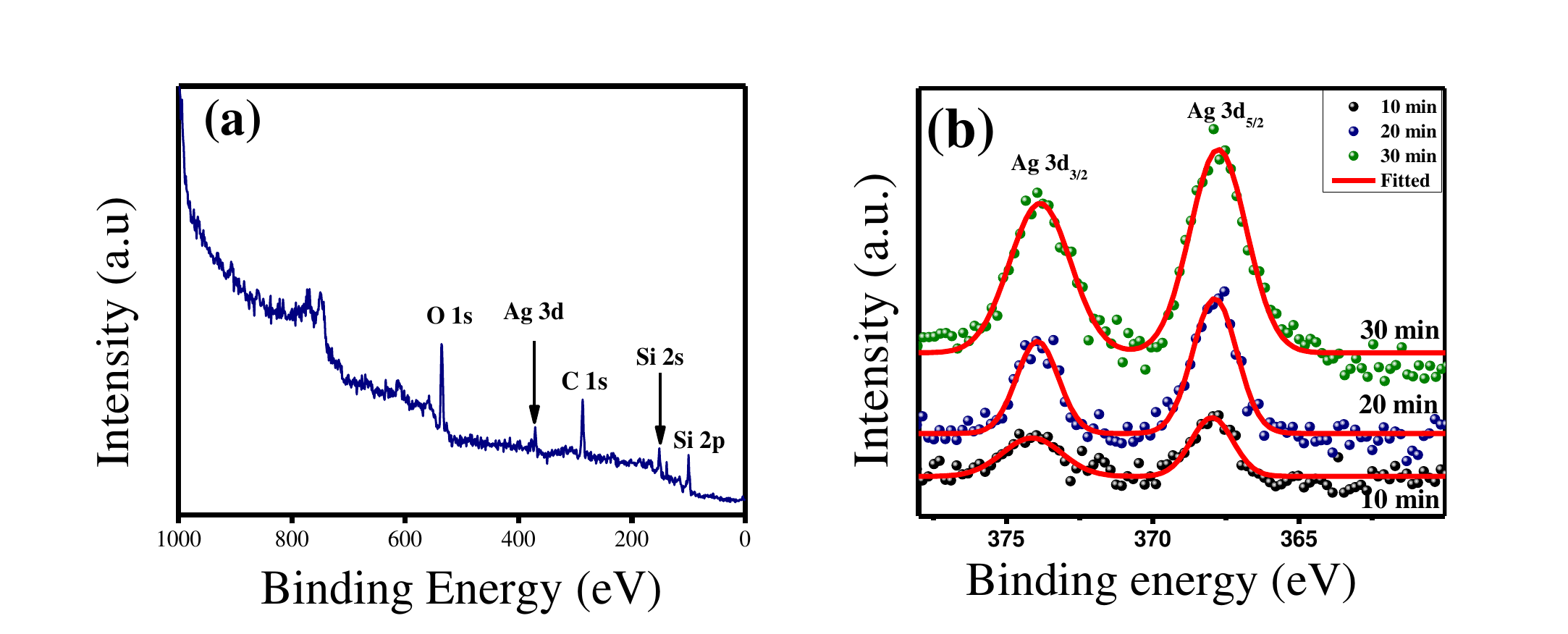}
\caption{\small (a) XPS survey spectrum for 30 min deposited film, (b) High resolution XPS spectrum of 30 min, 20 min and 10 min deposited sample}
\label{fig:fig_5}
\end{figure}

\begin{figure}[htb]
\centering
\captionsetup{width=.8\linewidth}
\includegraphics[scale=.65]{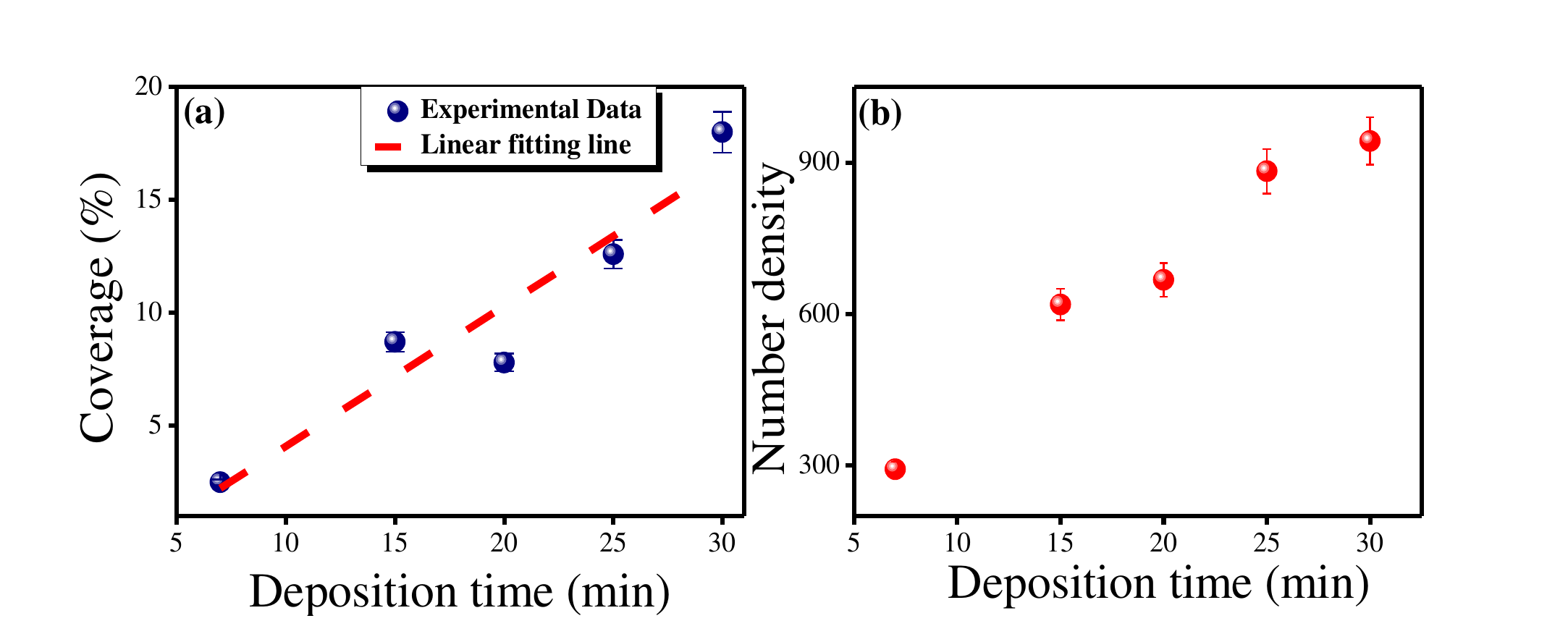}
\caption{\small (a) Coverage (in \%) and (b) number density calculated from the corresponding SEM images for different deposition time}
\label{fig:fig_7}
\end{figure}

The elemental composition of the deposited film was confirmed by the XPS measurement shown in \Figref{fig_5}a. From the full XPS spectrum, we confirm the presence of silver in our sample. For details analyses, we performed high resolution XPS spectra for the samples having deposition time 30 min, 20 min and 10 min. \Figref{fig_5}b shows the fitted Ag 3d core level XPS spectra of the samples. Binding energies of Ag 3d core level electrons are found as 368 eV,367.9 eV and 367.75 eV for deposition time of 10 min, 20 min and 30 min respectively. We also observe that with the increase of deposition time, amount of silver nanoparticle on the substrate increases which is reflected from the increase in peak intensity. The observed negative binding energy shift for higher deposition time indicates  the formation of large cluster-islands for higher deposition time, which is in agreement with the corresponding SEM and AFM images. In earlier studies, researchers found that the main  reason behind the binding energy shifting is due to charging effect on the surface\cite{ref17a}, size effect\cite{ref17b} and chemical effects\cite{ref17c}. H.S. Shin et al.\cite{ref17d} observed both the chemical as well as size effects of Ag nanoparticles and found that the negative binding energy shift is only due to larger particles. In our case, the only probable reason for binding energy shift is due to size effect as we can completly ignore the surface charge effect as well as the chemical effect on the surface.\\
\begin{figure}[htb]
\centering
\captionsetup{width=.8\linewidth}
\includegraphics[scale=.50]{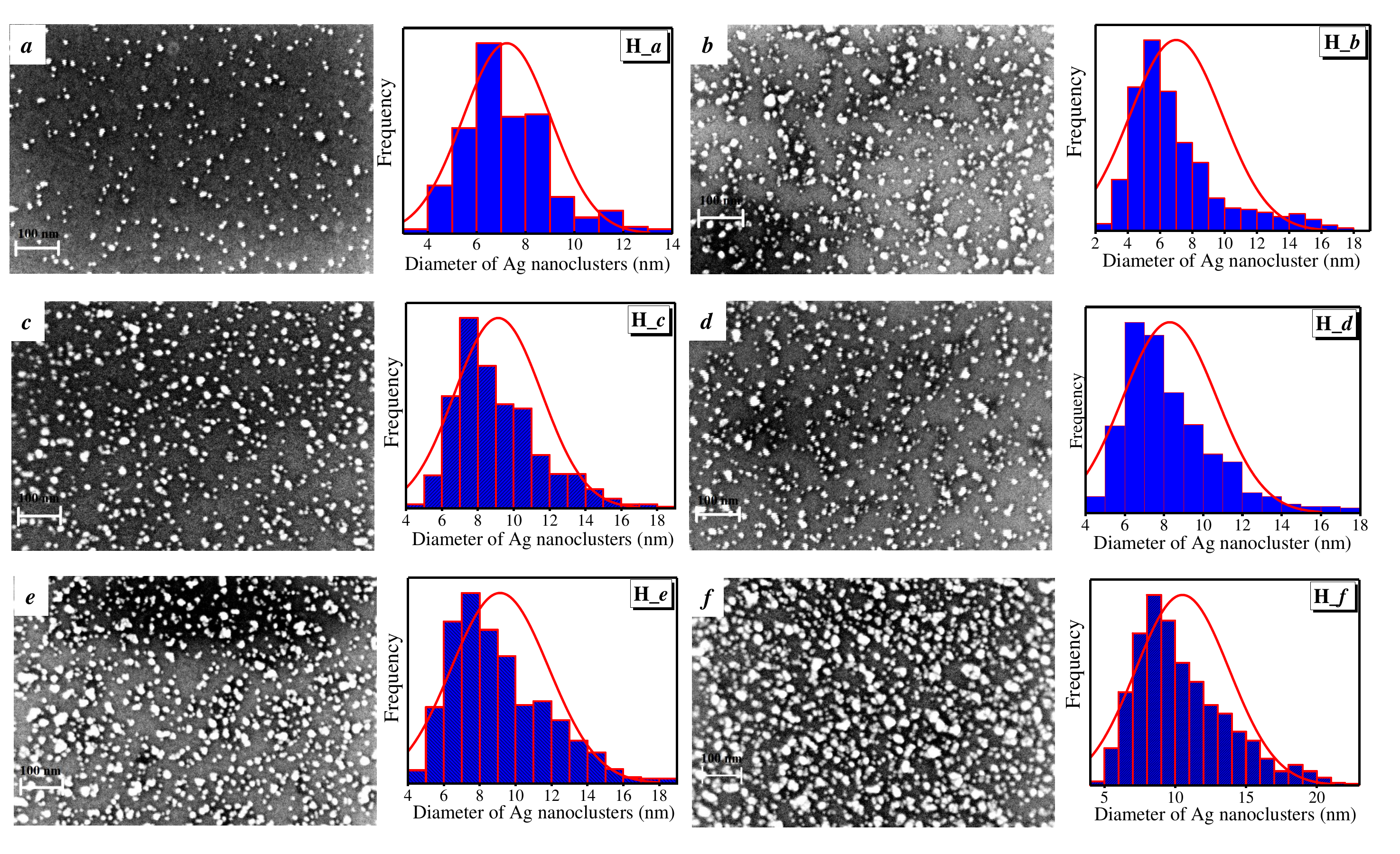}
\caption{\small SEM images of Ag nanoclusters deposited for \textit{a}) 7 min, \textit{b}) 10 min, \textit{c}) 15 min, \textit{d}) 20 min, \textit{e}) 25 min and \textit{f}) 30 min and the corresponding histograms for images \textit{a},\textit{b},\textit{c}, \textit{d}, \textit{e}, \textit{f} are H\_\textit{a}, H\_\textit{b}, H\_\textit{c}, H\_\textit{d}, H\_\textit{e}, H\_\textit{f}  }
\label{fig:fig_6}
\end{figure}
Moreover, we also examine  the surface morphology of the deposited Ag nanoclusters on the substrates for all the samples by means of Scanning Electron Microscopy (SEM) and are represented in \Figref{fig_6}. It is seen that as long as the deposited time increases, particle density on the surface becomes denser. Also from the corresponding size distributions of the deposited Ag nanoclusters for different deposition time shown in right side of \Figref{fig_6}, it is seen that the increased deposition time did not affect much on the mean cluster size of the deposited Ag nanoclusters. Although the size distributions of the deposited nanoclusters broaden towards higher values with the increase of deposition time. It is also noticed that for higher deposition time, number of bigger islands is more compared to that of the lower deposition time. It could be due to the aggregation/coalescences of Ag nanoclusters on the substrate surface. From these SEM images, we can also determine the number density of deposited Ag nanoclusters and their coverage of the surface. In \Figref{fig_7}(a), it is observed that the surface coverage value increases linearly upto a certain value with the increase of deposition time. The number density variation initial follows the trend of coverage variation with deposition time, but
later on at higher deposition time the rate is slow and tends to
saturate as shown in \Figref{fig_7}(b).

Summarizing the above results, we found two types of growth regimes of the silver films deposited by soft-landing deposition process. The value of  $\beta$ for the two regimes are found to be $0.27\pm0.07$ and $0.88\pm0.28$ for $t<20$ min and $t>20$ min respectively while the roughness exponent is found to be constant $\alpha=0.76\pm0.02$ for all deposition time. In many literature reviews, this type of scaling behavior has been observed. Buzio et al \cite{ref18} found self-affine surface of cluster assembled carbon film deposited on Si and Cu substrates at room temperature using a pulsed microplasma source. They found the roughness exponent $\alpha=0.64-0.68$ and growth exponent $\beta=0.42-0.52$ and concluded that these exponents were not influenced by different cluster size but only affected by the presence of large particle in the cluster beam. Some growth dynamic models are there which can properly explain the mechanism during surface growth. One of such model is Kardar-Parisi-Zhang (KPZ) model, in which the dominant relaxation mechanisms are desorption and/or vacancy formation on the surface \cite{ref10}. According to this model, the value of roughness exponent $\alpha_{KPZ}= 0.387$ and growth exponent $\beta_{KPZ}=0.25$. Palasantzas et all \cite{ref19,ref20} found that their experimental  roughness exponent $\alpha =0.45\pm0.05$ which is close to $\alpha_{KPZ}$, but their experimental growth exponent value ($\beta$) is very much deviated from the KPZ growth exponent value for low energy Cu nanocluster deposited on Si substrates. Since KPZ model did not consider surface diffusion as relaxation mechanism, the  surface diffusion occurs cannot be explain by this model\cite{ref10}. When surface diffusion is dominant over desorption or vacancy formation for relaxation mechanism, the Molecular Beam Epitaxy (MBE) model  \cite{ref11} comes into play. In this model the observed value of scaling exponents are $\alpha_{MBE} =0.67$ and $\beta_{MBE}= 0.2.$\\
From the discussions of the above models, we see that our experimental first growth region with a growth exponent value $\beta_1=0.26\pm0.07$ follows the KPZ growth exponent value $\beta_{KPZ}$, but the roughness exponent value does not agree with that of the KPZ roughness exponent. A similar case is reported by Jeffries et al.\cite{ref20a} for Pt film deposited on glass substrates at room temperature using sputtering process. They found their experimental value of roughness exponent as $\alpha\approx0.9\pm0.02$ and growth exponent as $\beta\approx0.26\pm0.03$, which they attributed to the linear diffusion. However the probability of cluster diffusion on the surface at room temperature is expected to be insignificant in the present work. So for the first growth regime, interface growth of size-selected Ag nanoclusters can be compared with KPZ growth exponenet.
\\
In the second growth regime associated with higher deposition time ($t>20$ min), the growth exponent value is found $\beta= 0.88\pm0.7$, which indicates very rough surface. Similar to our experimental value,  Yang and Xu observed the global surface fluctuation of polycrystalline Cu film at 700 K and their experimental growth exponent is $\beta=0.88$. Due to higher substrate temperature they explained the surface growth with the help of bulk diffusion process \cite{ref21}. In the present case, we can completely deny the bulk diffusion process as the experiment was performed at room temperature. Since none of the well-established theory of scaling can describe the obtained experimental results for the second growth regime, we can think of other non-local effect such as shadowing effect. The shadowing effect mainly depends on the height factor of a surface and it is applicable only for random angular distribution of the incoming flux or tilting the substrate during deposition \cite{ref22,ref23}. In this process, the surface undulation blocks the incoming flux from reaching the lower lying areas of the surface and thus, the hills receives more incoming particles and the valleys get only fewer particles. This allows the taller surface to grow faster at the expense of valleys. As a result, the overall surface will become much coarse.
\begin{figure}[htb]
\centering
\includegraphics[scale=.62]{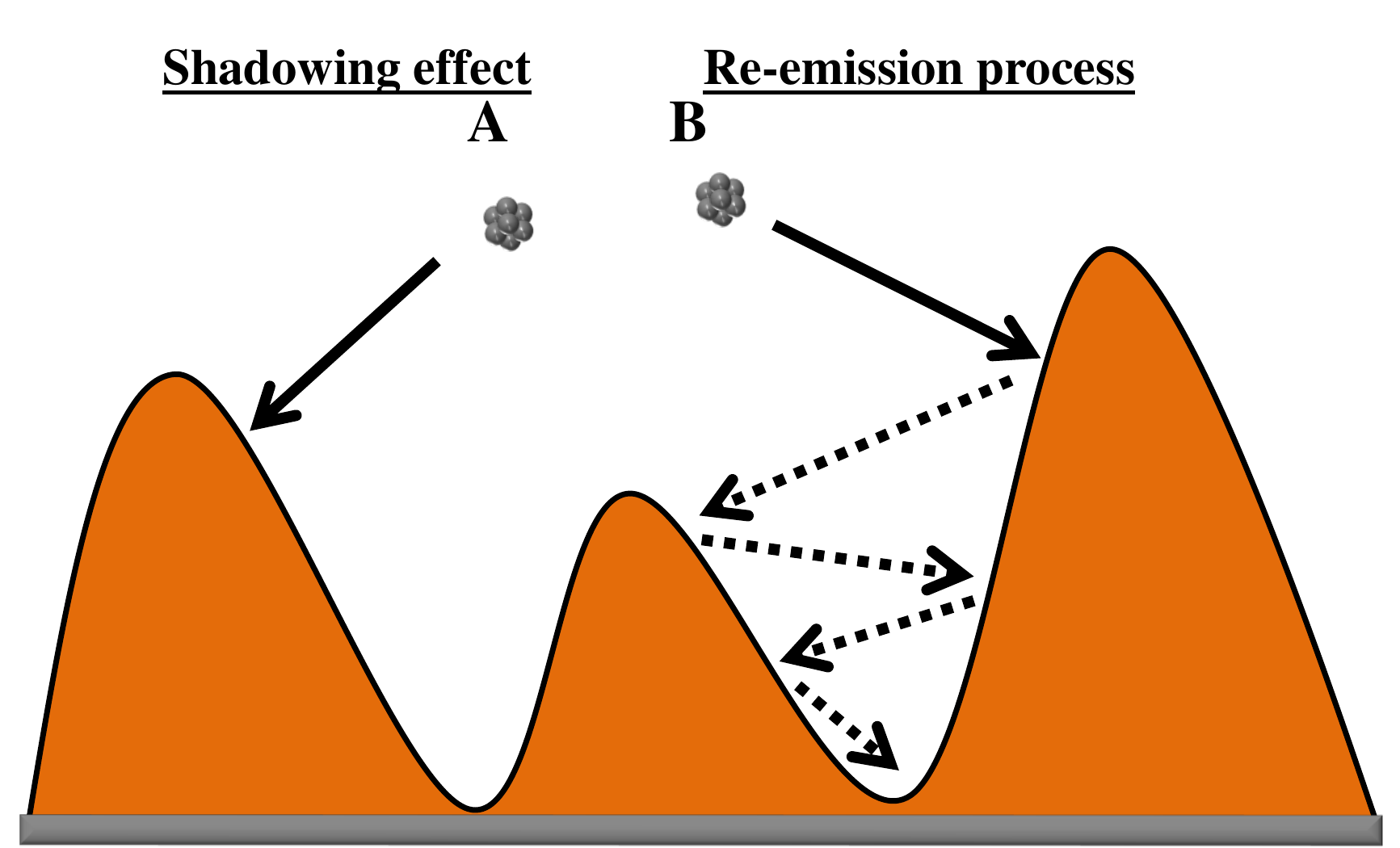}
\caption{\small Schematic illustration of shadowing and reemission effect}
\label{fig:fig_8}
\end{figure}
A schematic view of the shadowing process is given in \Figref{fig_8} according to the description by Karabacak et al \cite{ref24}. In the given diagram, A and B represent two of the incident clusters, where A is deposited on the taller surface due to shadowing effect while B is deposited on the valleys by following re-emission process. Here we can see that cluster A directly sticks on the taller surface, but B bounces back from its impact point and is re-emitted. This re-emitted cluster either sticks on the other surface point or again bounces off from the point; finally it gets deposited on the valleys. During angular deposition, shadowing effect increases the roughness of the surface but the reemission effect smoothens the surface. So, for a pure shadowing interface growth without any reemission, the growth exponent value is found to be equal 1, which is also obtained from the simulation result \cite{ref25,ref26}. In the present case, the growth of the size-selected Ag nanocluster film is found to be be dominated by the shadowing effect which leads to a growth exponent value $\beta=0.88\pm0.28$ close to 1.
Also, as clusters were deposited in soft landing process, the probability of occurring re-emission process is low. This is also a reason behind the dominance of growth exponent by shadowing process. The experiment was performed by tilting the substrate by $45^o$, as a result a non-uniform flux of clusters appeared on the surface. During deposition for shorter time, clusters are randomly deposited on the substrate which is found to follow the KPZ growth exponent. As the deposition time increases, the incoming clusters beam gets deposited on the already deposited clusters. Here these already deposited clusters acts as taller surface feature or hills. As the deposition time further increases, the hills will receive more clusters from the incoming cluster beams and form bigger cluster islands at the expense lower surface. This shadowing effect is also reflected from the height distribution graph shown in \Figref{fig_3}(d), where we observe that with increase of deposition time the height distribution becomes more bimodal in nature. From the particle size distribution, we get large cluster islands for larger deposition time, which also attributed to shadowing phenomena.

In the conclusion, we have reported the growth dynamics of size-selected silver nanoclusters deposited on silicon (100) substrates at room temperature using inert gas condensation technique with a dc magnetron. It is found that, the deposited size-selected Ag clusters shows two types of growth regimes and the surface morphology is self-affine in nature. Two growth regimes are separated by a cross over time of 20 min. In the first growth regime, surface roughness is found very low and the growth exponent value is found close to the KPZ growth model. For the second growth region, shadowing effect is found predominant for the growth of size-selected Ag film. The roughness exponent is found to be constant during all the regimes. The bimodal height distributions of the samples for higher deposition time ($t>20$ min) prove the shadowing phenomena that affects the surface morphology. Moreover the average reflectance of the surface increases with the increase of surface growth.
\section{Acknowledgement}
Two of the authors (P.B and A.D) would like to thank UGC for their financial support. We thank Prof. T.K. Chini for providing SEM facility. The authors also thankfully acknowledge Prof. P.M.G. Nambissan for providing UV-spectrometry facility and would like to thank Mrs Soma Roy for performing UV-spectroscopy. Finally authors thank Mr. Debraj Dey for technical support for smoothly conducting the experiment.


\begin{thebibliography}{10}

\bibitem{ref1} P. Meakin, Phys. Rep. \textbf{235}, 189 (1993).
\bibitem{ref2} T.H. Healy, Y.C. Zhang, Phys. Rep. \textbf{254}, 215 (1995).
\bibitem{ref3} J. Krug, Adv. Phys. \textbf{46}, 139 (1997).
\bibitem{ref4} J. Krug, in Scale Invariance, Interfaces, and Non-Equilibrium Dynamics, edited by A. McKane et al. (Plenum, New York, 1995).
\bibitem{ref5} T. Karabacak, Y.P. Zhao, G.C. Wang, T.M. Lu, Phys. Rev. B \textbf{66}, 075329 (2002).
\bibitem{ref6} P. Jensen, A.L. Barabasi, H. Larralde, S. Havlin, H.H. Stanley, Phys. Rev.  B \textbf{50}, 15316 (1994).
\bibitem{ref7} F. Family and T. Vicsek, Dynamics of Fractal Surfaces (World Scientific, Singapore, 1991).
\bibitem{ref8} A.L. Barabasi. and Stanley H. E., in Fractal Concepts in Surface Growth (Cambridge University Press,Cambridge, UK, 1995).
\bibitem{ref9} S.F. Edwards, D.R. Wikinson, Proc. R. Soc. Lond. A \textbf{381}, 17 (1982).
\bibitem{ref10} M. Kardar, G. Parisi, Y.C. Zhang, Phys. Rev. Lett.  \textbf{56}, 889 (1986).
\bibitem{ref11} S.D. Sarma, C.J. Lanczycki, R. Kotlyar, S.V. Ghaisas, Phys. Rev. E \textbf{53}, 359 (1996).
\bibitem{ref12} S. Mondal, D. Chowdhury, P. Barman, S.R. Bhattacharyya, Eur. Phys. J. D \textbf{71}, 327 (2017).
\bibitem{ref12a} Z.J. Jiang, C.Y. Liu, L.W. Sun, Journal of
Physical Chemistry  \textbf{109},  1730 ( 2005).
\bibitem{ref12b} M. Guzman, J. Dille, S. Godet, Nanomedicine  \textbf{8},  37 (2012).
\bibitem{ref12c} T.H. Lee, C.R. Hladik, R.M. Dickson, Appl Phys. Lett. \textbf{84}, 118 (2004).
\bibitem{ref12d} C. An, S. Wang, Y.Sun, Q. Zhang, J. Zhang, C. Wang, J. Fang, J. Mat. Chem. A  \textbf{4}, 4336 (2016).
\bibitem{ref12e} F.J. Beck, A. Polman, K.R. Catchpole, J. Appl. Phys. \textbf{105}, 114310 (2009).
\bibitem{ref12f} E.C. Tyo, S. Vajda, Nanture Nanotechnology \textbf{10}, 577 (2015).
\bibitem{ref12g} H. Tsunoyama, Y. Yamano, C. Zhang, M. Komari, T. Eguchi, A. Nakajima, Topics in Catalysis \textbf{61}, 126 (2018).
\bibitem{ref12h} K. Said, A.I. Ayesh, N.N. Qamhieh, F. Awwad, S.T. Mahmoud, S. Hisaindee, J. Alloys. Compd \textbf{694}, 1061 (2017).
\bibitem{ref12i} D. Biswas, S. Mondal, A. Rakshit, A. Bose, S.R. Bhattacharyya, S. Chakraborty, Mat. Sci. Semicon. Proc. \textbf{63}, 1 (2017).

\bibitem{ref13} F. Family, T. Vicsek, J. Phys. A \textbf{18}, L75 (1985).
\bibitem{ref14} Y.P. Zhao, G.C. Wang and  T.M. Lu, Characterization of Amorphous and Crystalline Rough Surfaces: Principles and Applications (Academic Press, San Diego, 2000).
\bibitem{ref15} D. Bhowmik, D. Chowdhury, P. Karmakar, Surf. Sci. \textbf{679}, 86 (2019).
\bibitem{ref16}  J. J. Ramasco, J. M. López,  M. A. Rodríguez, Phys. Rev. Lett. \textbf{84}, 2199 (2000).
\bibitem{ref17} S. Mondal, S. R. Bhattacharyya, Rev. Sci. Instrum. \textbf{85}, 065109 (2014).
\bibitem{ref17a} D.N.E Buchanan, Phys. Rev. B \textbf{33}, 5384 (1986).
\bibitem{ref17b} V. Vijayakrishnan, A. Chainani,D.D. Sarma, C.N.R. Rao, J. Phys. Chem \textbf{96} 8679 (1992).
\bibitem{ref17c} H.H. Huang, X.P. Ni, G.L. Loy, C.H. Chew, K.L. Tan, F.C. Loh, J.F. Deng, G.Q. Xu, Langmuir \textbf{12}, 909 (1996).
\bibitem{ref17d} H.S. Shin, H.C. Choi, Y. Jung, S.B. Kim, H.J. Song, H.J. Shin, Chem. Phys. Lett. \textbf{383}, 418 (2004).

\bibitem{ref18} R. Buzio, , E. Gnecco, C. Boragno, U. Valbusa, P. Piseri, E. Barborini, P. Milani, Surf. Sci. \textbf{444}, L1 (2000).
\bibitem{ref19} G.  Palasantzas, S. A. Koch, J. T. M. De Hosson,  Appl. Phys. Lett. \textbf{81}, 1089 (2002).
\bibitem{ref20} G. Palasantzas, S. A. Koch, J. T. M. De Hosson, Rev. Adv. Mater. Sci. \textbf{5}, 57 (2003).
\bibitem{ref20a} J.H. Jeffries, J.K. Zuo, M.M. Craig, Phys. Rev. Lett. \textbf{76}, 4931 (1996).
\bibitem{ref21} J.J. Yang, K.W. Xu, J. Appl. Phys. \textbf{101}, 104902 (2007).
\bibitem{ref22} T. Karabacak, J. Nanophotonics \textbf{5}, 052501 (2011).
\bibitem{ref23} J.H. Yao, H. Guo, Phys. Rev. E \textbf{ 47}, 1007 (1993).
\bibitem{ref24} T. Karabacak, H. Guclu, M. Yuksel, Phys. Rev. B \textbf{79}, 195418 (2009).
\bibitem{ref25} J.T. Drotar, Y.P. Zhao, T.M. Lu, G.C. Wang, Phys. Rev. B  \textbf{62}, 2118 (2000).
\bibitem{ref26} M. Pelliccione, T. Karabacak, T.M. Lu, Phys. Rev. Lett. \textbf{96}, 146105 (2006).


\end{thebibliography}
\end{document}